\documentclass[showpacs,preprint]{revtex4}
\usepackage{amssymb}
\usepackage{amsmath}
\usepackage{graphicx}
\usepackage{ulem}
\usepackage[dvips]{color}

\input{tcilatex}
\begin{document}

\title{Gamma-radiation in non-Markovian Fermi systems}
\author{V.M. Kolomietz$^{1)}$ , S.V. Radionov$^{1)}$ and B.V. Reznychenko$%
^{2)}$}
\affiliation{$^{1)}$Institute for Nuclear Research, 03680 Kiev, Ukraine\\
$^{2)}$Taras Shevchenko National University of Kyiv, 01601 Kiev, Ukraine}

\begin{abstract}
The gamma-quanta emission is considered within the framework of the
non--Markovian kinetic theory. It is shown that the memory effects have a
strong influence on the spectral distribution of gamma-quanta in the case of
long-time relaxation regime. It is shown that the gamma-radiation can be
used as a probe for both the time-reversible hindrance force and the
dissipative friction caused by the memory integral.
\end{abstract}

\pacs{PACS numbers : 05.45.+b,02.50.+s,03.65.-w,03.65.Ge}
\maketitle

\section{Introduction}

The dynamics and the dissipative properties of the many body Fermi system
depend in many aspects on the dynamic distortion of the Fermi surface in
momentum space. As is well-known \cite{abkh59}, the presence of Fermi
surface distortion allows the description of so-called collisional mechanism
of relaxation and gives rise to the damping of collective motion. An
additional one-body mechanism of relaxation exists in the finite system
where the particles are placed into the external mean field. The origin of
this damping is the collision of the particles with moving potential wall
\cite{bekh61}. We will consider below both of them.

On the other hand relaxation of collective motion implies fluctuations in
the corresponding collective variables, as follows from the
fluctuation-dissipation theorem. Furthermore, the fluctuations in a particle
density imply an accelerated motion of charges inside the charged system
like a nucleus and lead, therefore, to radiation. The spectral distribution
of this fluctuational radiation depends on the relaxation (dissipation)
properties of the collective motion, in particular, on the dynamic
distortion of the Fermi surface. We therefore suggest that a study of the
shape of the radiation spectrum emitted from the heated system provides an
opportunity to obtain information on the effects of temperature on
dissipative properties and on the transition from the low-temperature
(quantum) to the high-temperature (classical) regime in a finite many body
system.

In the present paper, we are interested in the spectrum of fluctuations in
shape variables. The precise form of such spectra can be expected to depend
on the parameters of the model, such as the collision time, and, especially,
on the memory effects. Here, we want to study these dependencies as one step
to our ultimate goal of determining the model parameters from a comparison
with experimental data, as might be possible due to a relation of the above
mentioned spectra to $\gamma $-spectra.

In what follows, we combine the thermal and quantum fluctuations by means of
the fluctuation-dissipation theorem. Such an approach presents a convenient
connection between different regimes of collective motion such as the
quantum zero-sound regime at zero temperature and the collisional
first-sound regime in a hot system. Such an approach presents a convenient
connection between different regimes of radiation such as the quantum regime
at zero's temperature and the thermal black body radiation of a hot system.

This paper is organized as follows. In Sec. II we suggest a proof of the
Langevin equation for the macroscopic collective variables starting from the
collisional Landau-Vlasov kinetic equation, including the memory effects in
the collisional integral. In Sec. III we review the classical approach to
the fluctuational radiation. We adopt a Langevin equation with a random
force as a source of the fluctuations. The main features of the dynamic
distortion of the Fermi surface are taken into account. In Sec. IV we apply
the results of Sec. II to the analysis of the spectral density of the
fluctuational radiation. Concluding remarks are presented in Sec. V.

\section{Surface fluctuations in a finite Fermi system}

To consider the fluctuations which accompany the collective motion in
many-body Fermi-system, one can start from the collisional kinetic equation
in presence of a random perturbation $y$ \cite{abkh59,kosh04}
\begin{equation}
{\frac{\partial }{\partial t}}f+{\frac{\mathbf{p}}{m}}\cdot \mathbf{\nabla }%
_{r}f-\mathbf{\nabla }_{r}U\cdot \mathbf{\nabla }_{p}f=\mathrm{St}[f]+y,
\label{1}
\end{equation}%
where $f\equiv f(\mathbf{r,p;}t)$ is the phase-space distribution function, $%
U\equiv U(\mathbf{r,p;}t)$ is the selfconsistent mean field and $\mathrm{St}%
[f]$ is the collision integral. The momentum distribution is distorted
during the time evolution of the system and takes the following form
\begin{equation}
f(\mathbf{r,p;}t)=f_{\mathrm{eq}}(\mathbf{r,p)\ +\ }\delta f(\mathbf{r,p;}%
t)=f_{\mathrm{sph}}(\mathbf{r,p;}t)+\sum_{lm,l\geq 1}\delta f_{lm}(\mathbf{%
r,p;}t),  \label{f}
\end{equation}%
where $f_{\mathrm{sph}}(\mathbf{r,p;}t)$ describes the spherical
distribution in momentum space, $l$ is the multipolarity of the
Fermi-surface distortion, $\delta f_{lm}$ is the component of the $l,m$
multipolarity in $\mathbf{p}$-space of the variation $\delta f$ and $f_{%
\mathrm{eq}}(\mathbf{r,p)}$ is the equilibrium distribution function. We
point out that the traditional time dependent Thomas-Fermi (\textrm{TDTF})
approximation is obtained from Eq. (\ref{1}) if one takes the distribution
function $f(\mathbf{r,p;}t)$ in the following restricted form $f_{\mathrm{TF}%
}(\mathbf{r,p;}t)=f_{\mathrm{sph}}(\mathbf{r,p;}t)+\delta f_{l=1}(\mathbf{%
r,p;}t)$ instead of Eq. (\ref{f}), see Ref. \cite{kota81}. Below we will
extend the \textrm{TDTF} approximation taking into consideration the dynamic
Fermi surface distortion up to multipolarity $l=2$ only and assume
\begin{equation}
\delta f=-\left( {{\frac{\partial f}{{\partial \epsilon }}}}\right) _{%
\mathrm{eq}}\sum_{l,m}^{l=2}\delta f_{lm}(\mathbf{r},t)Y_{lm}(\hat{p}).
\label{2.2}
\end{equation}%
Here, $\epsilon $ is the single particle energy and $(\partial f/{\partial
\epsilon })_{\mathrm{eq}}\sim \delta (\epsilon -\epsilon _{F})$, where $%
\epsilon _{F}$ is the Fermi energy \cite{abkh59}. Below we will restrict
ourselves to the azimuthally symmetric case (longitudinal perturbation)
where $\delta f_{lm}$ is $m$-independent.

We will consider a linear response to the external random perturbation $y$.
The linearized kinetic equation (\ref{1}) is given by%
\begin{equation}
{\frac{\partial }{\partial t}}\delta f+\hat{L}\delta f=\mathrm{\delta St}%
[f]+y  \label{2.1}
\end{equation}%
where $\mathrm{\delta St}[f]$ is\ the\ collision integral linearized in $%
\delta f=f-f_{\mathrm{eq}}$ and the operator $\hat{L}$ represents the drift
term
\begin{equation*}
\hat{L}\delta f={\frac{\mathbf{p}}{m}}\cdot \mathbf{\nabla }_{r}\delta f-%
\mathbf{\nabla }_{r}U_{\mathrm{eq}}\cdot \mathbf{\nabla }_{p}\delta f-%
\mathbf{\nabla }_{r}\delta U\cdot \mathbf{\nabla }_{p}f_{\mathrm{eq}}.
\end{equation*}

The collision integral $\mathrm{\delta St}[f]$ depends on the transition
probability of the two-nucleon scattering with initial momenta $(\mathbf{p}%
_{1},\mathbf{p}_{2})$ \ and final momenta $(\mathbf{p}_{1}^{\prime },\mathbf{%
p}_{2}^{\prime })$. At low temperatures $T\ll \epsilon _{F}$ the momenta $(%
\mathbf{p}_{1},\mathbf{p}_{2})$\ and $(\mathbf{p}_{1}^{\prime },\mathbf{p}%
_{2}^{\prime })$ are localized near the Fermi surface and the relaxation
time approximation can be used, see Refs. \cite{abkh59,kosh04},
\begin{equation}
\mathrm{\delta St}[f]=-{\frac{1}{\tau }}\left. \delta f\right\vert _{l\geq
1},  \label{st1}
\end{equation}%
where $\tau $ is the collisional relaxation time. The notation $l\geq 1$
means that the perturbation $\delta f|_{l\geq 1}$ in the collision integral
includes only Fermi surface distortions with a multipolarity $l\geq 1$ in
order to conserve the particle number in the collision processes \cite%
{abkh59}. The inclusion of the $l=1$ harmonic in the collision integral of
Eq. (\ref{st1}), at variance with the isoscalar case \cite{kola97}, is due
to nonconservation of the isovector current, i.e. due to a collisional
friction force between counterstreaming neutron and proton flows. The
relaxation time $\tau $ depends on the temperature and contains, in the
general case, memory effects ($\omega $-dependence) \cite{kosh04}:
\begin{equation}
\tau \equiv \tau (\omega ,T)={\frac{4\,\pi ^{2}\,\beta \,\hbar }{{(\hbar
\,\omega )^{2}+\zeta T^{2}}}}  \label{tau}
\end{equation}%
where $\beta $ and $\zeta $ are constants which are derived by the in-medium
nucleon-nucleon scattering. Note that the well-known \ Landau's prescription
\cite{lp} assumes $\zeta =4\,\pi ^{2}$. The parameter $\beta $ in Eq. (\ref%
{tau}) is rather badly established. It depends mainly on the in-medium
nucleon-nucleon scattering cross-section $\sigma _{NN}$. \ For example, this
value was calculated in Refs. \cite{kohl82} and \cite{wegm74} with the
results between $\beta =2.4$ and $\beta =19.3$ for different assumptions
about the scattering cross-section $\sigma _{NN}$.

Evaluating the first three moments of Eq. (\ref{2.1}) in $\mathbf{p}$-space
and taking into account the condition (\ref{2.2}), we can derive a closed
set of equations for the following moments of the distribution function,
namely, local particle density $\rho $, velocity field $u_{\nu }$ and
pressure tensor $P_{\nu \mu }$, in the form the continuity and Euler-like
equations (for details, see \textrm{Appendix }and Refs. \cite{kota81,kosh04}%
). We will restrict ourselves by the shape fluctuations of Fermi liquid
assuming an incompressible and irrotational flow, i.e.,

\begin{equation}
\nabla _{\nu }u_{\nu }=0  \label{2.15}
\end{equation}%
and assuming also a sharp particle distribution in $\mathbf{r}$-space

\begin{equation}
\rho =\rho _{0}\Theta \left[ R(t)-r\right] .  \label{2.16}
\end{equation}%
For the description of small amplitude oscillation of a certain
multipolarity $L$ we specify the surface as

\begin{equation}
r=R(t)=R_{0}\left[ 1+\sum_{M}\alpha _{LM}(t)Y_{LM}(\theta ,\phi )\right] .
\label{2.17}
\end{equation}%
The basic continuity and Euler-like equations can be then reduced to the
following Langevin equation (see \textrm{Appendix}, Eq. (\ref{2.28}))

\begin{equation}
-\omega ^{2}m_{L}\alpha _{LM,\omega }+(C_{L}^{(LD)}+C_{L}^{\prime }(\omega
))\alpha _{LM,\omega }-i\omega \gamma _{L}(\omega )\alpha _{LM,\omega }=\xi
_{LM,\omega },  \label{2.20}
\end{equation}%
where the index $\omega $ means the the Fourier transformation for the
corresponding values and $\xi _{LM,\omega }$ is the random force which
occurs due to the random perturbation $y$ in Eq. (\ref{1}). The left part of
Eq. (\ref{2.20}) derives the eigenfrequency of surface eigenvibrations of
the incompressible Ferm-liquid drop. Namely the corresponding secular
equation reads

\begin{equation}
-\omega ^{2}m_{L}+C_{L}^{(LD)}+C_{L}^{\prime }(\omega )-i\omega \gamma
_{L}(\omega )=0  \label{eigen1}
\end{equation}

In Eq. (\ref{2.20}), the mass coefficient $m_{L}$ is given by
\begin{equation}
m_{L}=m\int d\mathbf{r}\rho _{\mathrm{eq}}\sum_{\nu }|a_{LM,\nu }|^{2}={%
\frac{3}{4\pi L}}AmR_{0}^{2}  \label{2.21}
\end{equation}%
and the static stiffness coefficient $C_{L}^{(LD)}$ is derived from the
elastic properties of system

\begin{equation}
C_{L}^{(LD)}={\frac{1}{4\pi }}(L-1)(L+2)b_{S}A^{2/3}-\frac{L-1}{2L+1}b_{C}{%
\frac{Z^{2}}{A^{1/3}}},  \label{2.22}
\end{equation}%
where $b_{S}$ is the surface energy coefficient appearing in the nuclear
mass formula. This definition coincides with the one for the stiffness
coefficient in the traditional liquid drop model for the nucleus \cite{bomo2}%
. We point out, that the nucleon-nucleon interaction, manifested at the
starting equations (\ref{2.1}) and (\ref{2.2}), is presented in Eq. (\ref%
{2.20}) only implicitly through the phenomenological stiffness coefficient $%
C_{L}^{(LD)}$. Both coefficients $b_{S}$ and $b_{C}$ in Eq. (\ref{2.22}) are
temperature dependent. We will bellow assume the following temperature
dependence of the surface and Coulomb parameters \cite{rape83}
\begin{equation}
b_{S}=17.2\,\left[ {\frac{{T_{C}^{2}-T^{2}}}{{T_{C}^{2}+T^{2}}}}\right]
^{5/4}\,\mathrm{MeV},\,\,\,\,\,b_{C}={\frac{3}{2\pi }\frac{e^{2}}{r_{C}}}%
(1-x_{C}T^{2})\approx 0.55(1-x_{C}T^{2})\,\mathrm{MeV},  \label{bsbc}
\end{equation}%
where $r_{C}=1.24\ \mathrm{fm}$ \cite{bomo2}, the parameter $x_{C}$ was
chosen as $x_{C}=0.76\cdot 10^{-3}\,\mathrm{MeV}^{-2}$ and $T_{C}=18\,%
\mathrm{MeV}$ is taken as the critical temperature $T_{C}$ for infinite
nuclear Fermi-liquid \cite{rape83}. The nuclear Fermi-liquid does not exist
for temperatures $T\geq T_{C}$. Using Eq. (\ref{2.22}), one can find a
limiting temperature $T_{\mathrm{\lim }}^{(LD)}$ where the liquid drop
contribution $C_{L}^{(LD)}$ to the stiffness coefficient vanishes:
\begin{equation}
C_{L}^{(LD)}\equiv \left. C_{L}^{(LD)}(T)\right\vert _{T=T_{\mathrm{\lim }%
}^{(LD)}}=0.  \label{tlim}
\end{equation}%
For the parameters used in the present work one obtains $T_{\mathrm{\lim }%
}^{(LD)}=7.7\,\ \mathrm{MeV}$ for quadrupole deformation, $L=2$, in $^{208}%
\mathrm{Pb}$. For temperatures $T\geq T_{\mathrm{\lim }}^{(LD)}$ there
exists always an eigenfrequency with a positive imaginary part giving rise
to an exponentially growing deformation.

The additional stiffness coefficient $C_{L}^{\prime }(\omega )$ in Eq. (\ref%
{2.20}) is due to the Fermi surface distortion. In the case of quadrupole
distortions of Fermi surface, the final form of this coefficient is given by

\begin{equation}
C_{L}^{\prime }(\omega )=d_{L}\,{\frac{(\omega \tau )^{2}}{{1+(\omega \tau
)^{2}}}}\,P_{\mathrm{eq}},  \label{2.25}
\end{equation}%
where

\begin{equation*}
d_{L}=2{\frac{{(L-1)(2L+1)}}{L}}R_{0}^{3},\quad P_{\mathrm{eq}}=\frac{1}{3m}%
\ \int {\frac{gd\mathbf{p}}{(2\pi \hbar )^{3}}}p^{2}f_{\mathrm{eq}}={\frac{2%
}{5}\epsilon }_{F}\rho _{\mathrm{eq}}.
\end{equation*}%
Both stiffness coefficients $C_{L}^{(LD)}$ and $C_{L}^{\prime }(\omega )$
generate the shape eigenvibrations. The eigenfrequency $\omega _{L}$ of the
undamped (i.e., for $\gamma _{L}(\omega )=0$) eigenvibrations of
Fermi-liquid drop is obtained from the implicit equation
\begin{equation}
\omega _{L}(\omega )=\sqrt{\left[ C_{L}^{(LD)}+C_{L}^{\prime }(\omega )%
\right] /m_{L}}.  \label{2.26}
\end{equation}%
The friction coefficient $\gamma _{L}(\omega )$ in Eq. (\ref{2.20}) is given
by
\begin{equation}
\gamma _{L}(\omega )=d_{L}\,\eta (\omega ),  \label{2.27}
\end{equation}%
where $\eta (\omega )$ is the viscosity coefficient

\begin{equation}
\eta (\omega )={\frac{{\tau P}_{\mathrm{eq}}}{{1+(\omega \tau )^{2}}}}
\label{2.7}
\end{equation}

The secular equation (\ref{eigen1}) can be used to describe the eigenenergy $%
E$ and the width $\Gamma $ of the Giant Multipole Resonances (\textrm{GMR})
in cold nuclei. In \textrm{Fig. 1} and \textrm{2} we show the results of
calculations and the comparison with experimental data for the case of the
isoscalar Giant Quadrupole Resonances (\textrm{GQR}) for the nuclei through
the periodic table of elements. The numerical results in \textrm{Fig. 1} and
\textrm{2 }have been obtained using Eq. (\ref{eigen1}) and the relaxation
parameter $\beta =0.8\ \mathrm{MeV}$ in Eq. (\ref{tau}) for $T=0$.
Evaluating the \textrm{LDM} stiffness coefficient $C_{L}^{(LD)}$ of Eq. (\ref%
{2.22}), we have used the charge number $Z$ on the beta-stability which is
given by \cite{gren53}
\begin{equation*}
Z=\frac{1}{2}A\left[ 1-\frac{0.4\ A}{A+200}\right] .
\end{equation*}

\bigskip

As can be seen from \textrm{Fig. 1} and \textrm{2}, our approach provides a
quite satisfactory description of both the eigenenergies $E_{\mathrm{GQR}}$
and the widths $\Gamma _{\mathrm{GQR}}$ simultaneously. This fact can be
used to fit the relaxation parameter $\beta $ in Eq. (\ref{tau}). We will
below adopt \ $\beta =0.8\ \mathrm{MeV}$. We point out that the traditional
liquid drop model \cite{bomo2} is unable to describe the energy of the
\textrm{GQR}, see the dashed line in \textrm{Fig. 1}. In \textrm{Fig. 2 }the
dashed lines represent the results for the widths of the \textrm{GQR}
obtained by use of simplest Swiatecki's wall-formula (one-body dissipation)
given by \cite{nisi80,blbo78}
\begin{equation*}
\Gamma _{L,\mathrm{one-body}}=\frac{1}{A}\pi \rho _{0}v_{F}\hbar
R_{0}^{2}L\approx 34.3\ L\ A^{-1/3}\ \mathrm{MeV}
\end{equation*}%
and the so-called modified wall formula given by \cite{siko78}%
\begin{equation*}
\Gamma _{L,\mathrm{modif}}=\frac{1}{A}\pi \rho _{0}v_{F}\ \hbar \
\lambda ^{2}(L-1)^{2}L\approx 73.9\ (L-1)^{2}L\ A^{-1}\ \mathrm{MeV,}
\end{equation*}%
where the value of parameter $\lambda ^{2}$ $\approx 3\ \mathrm{fm}^{2}$ was
obtained from comparisons of calculated and experimental fission-fragment
kinetic energies. One can see that the simplest wall formula overestimates
significantly the \textrm{GQR} width while the calculation by use the
modified wall-formula is significantly smaller than the experimental results
for heavy nuclei.

Coming back to the right part of Eq. (\ref{2.20}), note that  the random
force $\xi _{LM,\omega }$ is derived by the properties for the ensemble
averaged correlation function. Namely,
\begin{equation}
\overline{\xi _{LM,\omega }}=0,\quad \overline{(\xi _{LM})_{\omega }^{2}}%
=2E(\omega ,T)\gamma _{L}(\omega ),  \label{2.29}
\end{equation}%
where, see also \cite{lali63},
\begin{equation}
E(\omega ,T)={\frac{\hbar \omega }{2}}\coth {\frac{\hbar \omega }{2T}=\ }%
\frac{\hbar \omega }{2}+\frac{\hbar \omega }{\exp \left( \hbar \omega
/T\right) -1}.  \label{2.14}
\end{equation}%
We have preserved the constant $\hbar $ in Eq. (\ref{2.14}) in order to
stress the fact that both the quantum and thermal fluctuations are involved
into the random force $\xi _{LM,\omega }$.

\section{Gamma-radiation caused by presence of random forces}

\bigskip

In this section we are going to establish the connection between the
spectrum of emitted photons and the equation of motion for the collective
variable, the fluctuations of which lead to the former of the radiation
spectrum. Let us start from the usual quantum-mechanical definition of the
perturbative transition probability per unit time in an energy interval $%
d(\hbar \omega )$
\begin{equation}
dW_{fi}=\frac{1}{\hbar ^{2}}\lim_{\mathcal{T}\rightarrow \infty }{\frac{1}{2%
\mathcal{T}}}\left\vert \int_{-\mathcal{T}}^{\mathcal{T}}dt^{\prime
}\left\langle \psi _{f}(t^{\prime })|V(t^{\prime })|\psi _{i}(t^{\prime
})\right\rangle \right\vert ^{2}d\nu _{f}.  \label{3.1}
\end{equation}%
Here $V(t)$ is the one-body perturbation field
\begin{equation}
V(t)=e^{i\omega t}\int d\mathbf{r}\ e\ q(\mathbf{r})\hat{\rho}(\mathbf{r})+%
\mathrm{c.c.},  \label{3.2}
\end{equation}%
where $e\hat{\rho}(\mathbf{r})$ is the charge density operator
\begin{equation}
e\hat{\rho}(\mathbf{r})=\sum_{i=1}^{A}e_{i}\delta (\mathbf{r}-\vec{r_{i}})
\label{3.3}
\end{equation}%
and $d\nu _{f}$ is the number of final states in the energy interval $[\hbar
\omega ,\hbar \omega +d(\hbar \omega )]$. The choice for the function $q(%
\mathbf{r})$ depends on the problem under consideration.

In a general case the initial and final wave functions, $\psi _{i}(t)$ and $%
\psi _{f}(t)$ respectively, are non-stationary ones and we write
\begin{equation}
\psi _{f}^{\star }(t)\psi _{i}(t)=\varphi _{f}^{\star }(\vec{r_{1}},...,\vec{%
r_{A}})\varphi _{i}(\vec{r_{1}},...,\vec{r_{A}})\alpha _{fi}(t).  \label{3.4}
\end{equation}%
In particular, in a stationary case we have
\begin{equation}
\alpha _{fi}=e^{i(E_{f}-E_{i})t/\hbar }.  \label{3.5}
\end{equation}%
and Eqs. (\ref{3.1}) and (\ref{3.2}) give the usual result for the
transition probability per unit time which is $dW_{fi}\sim \delta
(E_{i}-E_{f}-\hbar \omega )$.

We will consider below the electromagnetic $EL$-transitions using for $q(%
\mathbf{r})$ in (\ref{3.2}),
\begin{equation}
q(\mathbf{r})=r^{L}Y_{LM}\equiv q_{LM}(\mathbf{r}).  \label{3.6}
\end{equation}%
The usual transformation of Eq. (\ref{3.1}) for the case of multipole
transitions gives \cite{fisa90,beli82}
\begin{equation*}
dW_{fi}(EL)=4{\frac{L+1}{L}}\,{\frac{1}{[(2L+1)!!]^{2}}}\frac{e^{2}}{\hbar }%
k^{2L+1}{\frac{1}{{2J_{i}+1}}}\sum_{M_{i}M_{f}M}\left\vert \left\langle
\varphi _{f}\left\vert \int d\mathbf{r}q_{LM}(\mathbf{r})\hat{\rho}%
\right\vert \varphi _{i}\right\rangle \right\vert ^{2}
\end{equation*}%
\begin{equation}
\times \lim_{\mathcal{T}\rightarrow \infty }{\frac{1}{2\mathcal{T}}}%
\left\vert \int_{-\mathcal{T}}^{\mathcal{T}}dt^{\prime }e^{i\omega t^{\prime
}}\alpha _{fi}(t^{\prime })\right\vert ^{2}d\omega ,  \label{3.7}
\end{equation}%
where $kc=\omega $ is the wave number of the photon.

The transition probability $dW_{fi}$ allows us to evaluate the power $%
dP_{fi} $ radiated in the energy interval $d(\hbar \omega )$ as
\begin{equation}
dP_{fi}(EL)=\hbar \omega dW_{fi}(EL).  \label{3.8}
\end{equation}%
The classical result the radiated power $dP_{\mathrm{class}}(EL)$ can be
obtained from the quantum mechanical one, Eqs. (\ref{3.7}) and (\ref{3.8}),
by using the correspondence principle for the transition density:
\begin{equation}
\left\langle \psi _{f}(t)|e\hat{\rho}|\psi _{i}(t)\right\rangle \equiv
\alpha _{fi}(t)\left\langle \varphi _{f}|e\hat{\rho}|\varphi
_{i}\right\rangle \Rightarrow e\delta \rho (\mathbf{r},t)=\alpha (t)\
e\delta \rho (\mathbf{r})\,.  \label{3.9}
\end{equation}%
Here $e\delta \rho (\mathbf{r},t)$ is the variation of the classical charge
density in the external field $V(t)$, Eq. (\ref{3.2}). Thus, we have from
Eqs. (\ref{3.7})-(\ref{3.9})
\begin{equation}
dP_{\mathrm{class}}(EL)=R_{L}(\omega )\sum_{M}\left\vert Q_{LM}\right\vert
^{2}\lim_{\mathcal{T}\rightarrow \infty }{\frac{1}{2\mathcal{T}}}\left\vert
\int_{-\mathcal{T}}^{\mathcal{T}}dt^{\prime }e^{i\omega t^{\prime }}\alpha
(t^{\prime })\right\vert ^{2}d\omega ,  \label{3.10}
\end{equation}%
where
\begin{equation}
R_{L}(\omega )=4{\frac{L+1}{L[(2L+1)!!]^{2}}\omega }e^{2}\ \left( \frac{%
\omega }{c}\right) ^{2L+1}  \label{3.11}
\end{equation}%
and

\begin{equation}
Q_{LM}=\int d\mathbf{r}q_{LM}(\mathbf{r})\delta \rho (\mathbf{r}).
\label{3.12}
\end{equation}

Let us rewrite the time double integral in Eq. (\ref{3.10}) as%
\begin{equation*}
\lim_{\mathcal{T}\rightarrow \infty }{\frac{1}{2\mathcal{T}}}\left\vert
\int_{-\mathcal{T}}^{\mathcal{T}}dte^{i\omega t}\alpha (t)\right\vert
^{2}=\lim_{T\rightarrow \infty }{\frac{1}{2T}}\int_{-\mathcal{T}}^{\mathcal{T%
}}dt\int_{-\mathcal{T}}^{\mathcal{T}}dt^{\prime }e^{i\omega t^{\prime
}}\alpha (t)\alpha (t+t^{\prime })
\end{equation*}%
\begin{equation}
=\int_{-\infty }^{\infty }dt^{\prime }e^{i\omega t^{\prime }}\overline{%
\alpha (t)\alpha (t+t^{\prime })}.  \label{3.13}
\end{equation}%
Here the time average
\begin{equation}
\overline{\alpha (t)\alpha (t+t^{\prime })}=\lim_{\mathcal{T}\rightarrow
\infty }{\frac{1}{2\mathcal{T}}}\int_{-\mathcal{T}}^{\mathcal{T}}dt\alpha
(t)\alpha (t+t^{\prime })=\left\langle \alpha (t)\ \alpha (t+t^{\prime
})\right\rangle  \label{3.14}
\end{equation}%
can be considered as an ensemble average $\left\langle ...\right\rangle $
for an ergodic system.

Finally, we shall rewrite Eq. (\ref{3.10}) as
\begin{equation}
dP_{\mathrm{class}}(EL)=R_{L}(\omega )\sum_{M}Q_{LM}^{2}\left\langle \alpha
^{2}\right\rangle _{\omega }d\omega ,  \label{3.15}
\end{equation}%
where $\left\langle \alpha ^{2}\right\rangle _{\omega }$ is the spectral
correlation function\ \cite{lali63}
\begin{equation}
\left\langle \alpha ^{2}\right\rangle _{\omega }=\int_{-\infty }^{\infty
}dt^{\prime }e^{i\omega t^{\prime }}\left\langle \alpha (t)\ \alpha
(t+t^{\prime })\right\rangle .  \label{3.16}
\end{equation}

As it was demonstrated in the previous section, the dynamics of small
fluctuations of the collective variable $\alpha (t)$ coupled to a heat bath
can be described by a Langevin equation of the form
\begin{equation}
-m_{L}\omega ^{2}\alpha _{LM,\omega }-i\gamma _{L}(\omega )\omega \alpha
_{LM,\omega }+m_{L}\omega _{L}^{2}(\omega )\alpha _{LM,\omega }=\xi
_{LM,\omega }.  \label{3.17}
\end{equation}%
Using Eqs. (\ref{3.15}) and (\ref{3.17}) we can derive an expression for the
emitted fluctuational power in terms of the spectral correlation function $%
\overline{(\xi _{LM})_{\omega }^{2}}$ of the random force $\xi _{LM}(t)$:

\begin{equation}
dP_{\mathrm{class}}(EL)=R_{L}(\omega )\sum_{M}Q_{LM}^{2}{\frac{\overline{%
(\xi _{LM})_{\omega }^{2}}}{{m_{L}^{2}}\left[ {\omega ^{2}-\omega
_{L}^{2}(\omega )}\right] {^{2}+\omega ^{2}\gamma _{L}^{2}(\omega )}}}%
d\omega .  \label{3.18}
\end{equation}%
The function $\overline{(\xi _{LM})_{\omega }^{2}}$ depends on the
dissipative properties of the system. In the previous section we have
derived this function as well as explicit expressions for the transport
coefficients $m_{L},\omega _{L}$ and $\gamma _{L}$ for the case of
collective particle excitation, see Eq. (\ref{2.29}).

\section{Spectral density of radiation. Numerical results}

\bigskip

Equation (\ref{2.20}) for the shape oscillations of a Fermi-liquid drop
together with Eqs. (\ref{2.29}) and (\ref{3.18}) can be used for the
analysis of the spectral density
\begin{equation}
J_{L}(\omega )=dP_{\mathrm{class}}(EL)/d\omega   \label{4.1}
\end{equation}%
of fluctuational radiation. We thus have
\begin{equation}
J_{L}(\omega )=\sum_{M}R_{L}(\omega )Q_{LM}^{2}{\frac{\overline{(\xi
_{LM})_{\omega }^{2}}}{{m_{L}^{2}}\left[ {\omega ^{2}-\omega _{L}^{2}(\omega
)}\right] {^{2}+\omega ^{2}\gamma _{L}^{2}(\omega )}}}.  \label{4.2}
\end{equation}%
In the case of shape oscillations of $L$ multipolarity we have from Eqs. (%
\ref{3.9}), (\ref{2.16}) and (\ref{2.17})
\begin{equation}
\delta \rho (\mathbf{r},t)=-\rho _{0}R_{0}\delta (r-R_{0})\sum_{M}\alpha
_{LM}(t)Y_{LM}(\theta ,\phi )  \label{4.3}
\end{equation}%
and from Eqs. (\ref{3.6}), (\ref{3.9}) and (\ref{3.12})
\begin{equation}
Q_{LM}=-\rho _{0}R_{0}^{L+3}.  \label{4.4}
\end{equation}%
Collecting Eqs. (\ref{4.2}), (\ref{2.29}) and (\ref{4.4}) we finally\ find
\begin{equation*}
J_{L}(\omega )=H_{L}{\frac{E(\omega ,T)\gamma _{L}(\omega )}{{m_{L}^{2}}%
\left[ {\omega ^{2}-\omega _{L}^{2}(\omega )}\right] {^{2}+\omega ^{2}\gamma
_{L}^{2}(\omega )}}}
\end{equation*}%
\begin{equation}
=\left[ \frac{1}{2}\hbar \omega +\frac{\hbar \omega }{\exp \left( \hbar
\omega /T\right) -1}\right] {\frac{H_{L}\gamma _{L}(\omega )}{{m_{L}^{2}}%
\left[ {\omega ^{2}-\omega _{L}^{2}(\omega )}\right] {^{2}+\omega ^{2}\gamma
_{L}^{2}(\omega )}},}  \label{4.5}
\end{equation}%
where
\begin{equation}
H_{L}=8e^{2}\rho _{0}^{2}{\omega \ }\left( \frac{\omega }{c}\right) ^{2L+1}{%
\frac{{L+1}}{L}}{\frac{R_{0}^{2L+6}}{[(2L+1)!!]^{2}}}.  \label{4.6}
\end{equation}%
Note, that presence of term $\hbar \omega /2$ in Eq. (\ref{4.5}) reflects a
general problem of zero energy contribution. This term provides an
unphysical infinite contribution to the total energy of radiation and must
be thereby excluded. We preserve this term to provide the correct transition
to the quantum regime at $T\rightarrow 0$, where this term manifests the
zero-point fluctuations.

In a general case, we have to take into account the radiation friction
effects in Eq. (\ref{4.5}) to guarantee the asymptotic convergency of the
spectral density $J_{L}(\omega )$ at $\omega \rightarrow \infty $. It can be
done by the following substitution for $\gamma _{L}(\omega )$ in the
denominator of Eq. (\ref{4.5}) (see Ref. \cite{fisa90})
\begin{equation}
\gamma _{L}(\omega )\rightarrow \gamma _{L}(\omega )+\left( \omega /\omega
_{L}\right) ^{2}\gamma _{L}^{\prime }(\omega ).  \label{4.7}
\end{equation}%
Here $\gamma _{L}^{\prime }(\omega )$ is the radiation friction coefficient
\begin{equation}
\gamma _{L}^{\prime }(\omega )=\Gamma _{L}/\omega ,  \label{4.8}
\end{equation}%
$\Gamma _{L}$ is the radiation width of the surface excitation
\begin{equation}
\Gamma _{L}=(1/\omega )\pi H_{L}\alpha _{L,0}^{2}  \label{4.9}
\end{equation}%
and $\alpha _{L,0}$ is the zero-point amplitude
\begin{equation}
\alpha _{L,0}^{2}={\frac{\hbar }{{2\sqrt{C_{L}\,m_{L}}}}=\frac{\hbar }{{%
2\omega _{L}m_{L}}}}\,\,.  \label{4.10}
\end{equation}%
The formula (\ref{4.5}) for the radiation is valid for arbitrary collision
times $\tau $ and thus describes both the quantum and the high temperature
limit as well as the intermediate cases. From it one can obtain the leading
order terms in the different limits mentioned.

\textit{(1) High temperature limit:} $\omega \tau \rightarrow 0,T\gg \hbar
\omega $ \newline
The contribution from the dynamic distortion of the Fermi surface can be
neglected in this case and we have from Eq. (\ref{2.25})

\begin{equation}
C_{L}^{\prime }(\omega _{L})\approx 0.  \label{4.11}
\end{equation}%
The eigenfrequencies $\omega _{L}$ of the shape oscillations are determined
here by the usual liquid drop model as
\begin{equation}
\omega _{L,0}=\sqrt{C_{L}/m_{L}}\,\,.  \label{4.12}
\end{equation}%
In the high temperature regime, the Fermi liquid viscosity $\eta (\omega )$,
Eq. (\ref{2.7}), approaches the classical expression \cite{abkh59}
\begin{equation}
\eta _{\mathrm{class}}={\frac{1}{5}}\rho _{0}p_{F}^{2}\tau _{0},
\label{4.13}
\end{equation}%
where $p_{F}$ is the Fermi momentum and $\tau _{0}\equiv \tau (\omega =0,T)$%
. The spectral correlation function $\overline{(\xi _{LM})_{\omega }^{2}}$
of the random force can be found from Eqs. (\ref{2.29}), (\ref{2.14}) and (%
\ref{2.27})
\begin{equation}
\overline{(\xi _{LM})_{\omega }^{2}}=2\gamma _{L,0}T\,.  \label{4.14}
\end{equation}%
where $\gamma _{L,0}=\gamma _{L}(\omega =0)$. This correlation function is
independent of $\omega $, i.e., it corresponds to a white noise.

In this limit, the spectral density of radiation, Eq. (\ref{4.5}), is given
by

\begin{equation}
J_{L}(\omega )=H_{L}{\frac{\gamma _{L,0}T}{{m_{L}^{2}}\left( {\omega ^{2}-{%
\omega _{L,0}^{2}}}\right) {^{2}+\omega ^{2}}\gamma _{L,0}^{2}},}
\label{4.15}
\end{equation}%
where ${{\omega _{L,0}}}=\sqrt{C_{L}/m_{L}}$. The spectral density (\ref%
{4.15}) is proportional to the temperature $T$ as expected for a classical
thermal emission of radiation \cite{levi71}. In the high temperature limit $%
T\rightarrow \infty $ we have $\gamma _{L,0}\rightarrow 0$ and
\begin{equation}
\lim_{T\rightarrow \infty }{\frac{\gamma _{L,0}}{{m_{L}^{2}(\omega ^{2}-{%
\omega _{L,0}^{2}})+\omega ^{2}}\gamma _{L,0}^{2}}}={\frac{\pi }{2m_{L}{{%
\omega _{L,0}^{2}}}}}[\delta (\omega -{{\omega _{L,0}}})+\delta (\omega +{{%
\omega _{L,0}}})].  \label{4.16}
\end{equation}%
Thus, the spectral density of the radiation is given at high temperature by
\begin{equation}
J_{L}(\omega )=\pi H_{L}\alpha _{L,\mathrm{therm}}^{2}\delta (\omega -{{%
\omega _{L,0}}}),  \label{4.17}
\end{equation}%
where $\alpha _{L,\mathrm{therm}}^{2}$ is the square of the thermal
oscillation amplitude
\begin{equation}
\alpha _{L,\mathrm{therm}}^{2}=\frac{T}{2m_{L}{{\omega _{L,0}^{2}}}}=\frac{T%
}{2C_{L}}.  \label{4.18}
\end{equation}%
The result of Eqs. (\ref{4.17}) and (\ref{4.18}) recovers the
Rayleigh--Jeans law for the black body radiation.

\textit{(2) Quantum regime: }$\omega \tau \rightarrow \infty ,\,\,T\ll \hbar
\omega $ \newline
The contribution to the stiffness coefficient from the dynamic distortion of
the Fermi surface is now given by (see Eq. (\ref{2.25}))
\begin{equation}
C_{L}^{\prime }(\omega )\approx \widetilde{C}_{L}^{\prime }=d_{L}\,P_{%
\mathrm{eq}}.  \label{4.19}
\end{equation}%
We note that, in a cold Fermi system at $L\neq 1$, $\widetilde{C}%
_{L}^{\prime }$ provides the main contribution to the stiffness coefficient.
The viscosity coefficient $\eta (\omega )$, Eq. (\ref{2.7}), can be
approximated in this limit by

\begin{equation}
\eta (\omega )=(P_{\mathrm{eq}}/4\,\pi ^{2}\,\beta \,\hbar )[1+\zeta
(T/\omega )^{2}].  \label{4.20}
\end{equation}%
The spectral correlation function $\overline{(\xi _{LM})_{\omega }^{2}}$ is
obtained from Eqs. (\ref{2.29}), (\ref{2.27}) and (\ref{2.14}) to be
\begin{equation}
\overline{(\xi _{LM})_{\omega }^{2}}=\hbar \omega \widetilde{\gamma }_{L},
\label{4.21}
\end{equation}%
where
\begin{equation}
\widetilde{\gamma }_{L}=d_{L}\,P_{\mathrm{eq}}/4\,\pi ^{2}\,\beta \,\hbar
\label{4.22}
\end{equation}%
does not depend on $\omega $. The spectral correlation function (\ref{4.21})
now corresponds to a blue noise.

The spectral density of radiation $J_{L}(\omega )$ can be found from Eqs. (%
\ref{4.5}) and (\ref{4.21}) to have the form
\begin{equation}
J_{L}(\omega )=H_{L}{\frac{\hbar \omega \widetilde{\gamma }_{L}}{{%
m_{L}^{2}(\omega ^{2}-\widetilde{\omega }_{L}^{2})^{2}+\omega ^{2}\widetilde{%
\gamma }_{L}^{2}}}},  \label{4.23}
\end{equation}%
where $\widetilde{\omega }_{L}$ is the eigenfrequency
\begin{equation*}
\widetilde{\omega }_{L}=\sqrt{(C_{L}+\widetilde{C}_{L}^{\prime })/m_{L}}
\end{equation*}%
of the zero sound mode in the case of no damping.

Similarly to the high temperature regime, for low temperatures the spectral
density (\ref{4.23}) takes the form of a sharp peak in the limit of small
damping, i.e. for $\widetilde{\gamma }_{L}\rightarrow 0$:
\begin{equation}
J_{L}(\omega )=\lim_{\widetilde{\gamma }_{L}\rightarrow 0}H_{L}{\frac{\hbar
\omega \widetilde{\gamma }_{L}}{{m_{L}^{2}(\omega ^{2}-\ }\widetilde{{\omega
}}{{_{L}^{2}})^{2}+\omega ^{2}\widetilde{\gamma }_{L}^{2}}}}=H_{L}{\frac{\pi
\hbar }{2m_{L}\widetilde{\omega }_{L}}\delta (\omega -\widetilde{\omega }%
_{L})=\quad }\pi H_{L}\widetilde{\alpha }_{L,0}^{2}\delta (\omega -%
\widetilde{\omega }_{L}),  \label{4.24}
\end{equation}%
where (compare with Eq. (\ref{4.18}))%
\begin{equation*}
\widetilde{\alpha }_{L,0}^{2}=\frac{\hbar \widetilde{\omega }_{L}}{2m_{L}%
\widetilde{{\omega }}{{_{L}^{2}}}}=\frac{\hbar \widetilde{\omega }_{L}}{%
2(C_{L}+\widetilde{C}_{L}^{\prime })}
\end{equation*}%
is the square of renormalized zero-point amplitude (compare with Eq. (\ref%
{4.18})). In this limit the expression for the spectral density (\ref{4.24})
coincides with the usual quantum-mechanical result for the photon emission
associated with shape oscillations of the charge $Z$. We recall that the
quantum-mechanical result (\ref{4.24}) was obtained from the classical
approach, Eq. (\ref{3.18}). It is due to the fact that the quantum
fluctuations have been incorporated into the correlation function (\ref{2.29}%
) through the factor $E(\omega ,T)$, Eq. (\ref{2.14}), see also \cite{lali63}.

In \textrm{Fig. 3} we have plotted the spectral density of gamma-quanta
emission $J_{L}(\omega )$ as obtained from Eq. (\ref{4.5}) for two
temperatures $T=3\,\mathrm{MeV}<T_{\mathrm{\lim }}^{(LD)}$ and $T=8\,\mathrm{%
MeV}>T_{\mathrm{\lim }}^{(LD)}$ in the case $\beta =0.8\ \mathrm{MeV}$. The
dashed line is for the statistical $\gamma $-quanta emission given by \cite%
{BW}%
\begin{equation}
J_{L}(\omega )=\text{\textrm{const\ }}\omega ^{2L+1}\exp \left( -\frac{\hbar
\omega }{T}\right) ,  \label{BW}
\end{equation}%
where the value of "const" is normalized to the same integral emission as is
obtained from Eq. (\ref{4.5}). For low temperature $T=3\,\mathrm{MeV}$ we
observe a well defined maximum (solid curve 1) which corresponds to the
\textrm{GQR} excitation (zero-sound regime).

\bigskip

An increase of $T$ leads to a shift of the maximum of $J_{L}(\omega )$ to
lower frequencies and to an increase in the width. The shape of the curves
near the zero-sound maximum in \textrm{Fig. 3} is a non-Lorentzian one and
depends, in particular, on the retardation effects in the friction
coefficient, Eq. (\ref{2.27}), and, consequently, on the parameters $\beta $
and $\zeta $ in the relaxation time, Eq. (\ref{tau}). Increasing the
temperature we do not find a first sound peak centered at low frequency. \
We point out an interesting phenomenon. For a large enough value of $\beta $%
, namely $\beta \geq 0.5\,\mathrm{MeV}$, there is, in principle, a
possibility for a resonance-like structure of $J_{L}(\omega )$ at
temperatures $T>T_{\mathrm{\lim }}^{(LD)}$ which is due to the pure
Fermi-surface vibrations in the momentum space. For these values of $\beta $
there exists a temperature region where $C_{L}^{(LD)}(T)\leq 0$ but $%
C_{L}(\omega _{L})>0$, simultaneously. This implies the existence, in this
high temperature region, of a particular eigenmode of the Fermi liquid drop
where the restoring force is exclusively due to the dynamic Fermi-surface
distortion. Note that the non-monotonic behavior of solid curve 2 in \textrm{%
Fig. 3} occurs just due to the combination of polynomial $\sim \omega
^{2L+2} $ and the Planck's $\sim \left[ \exp \left( \hbar \omega /T\right) -1%
\right] ^{-1}$ multipliers in Eq. (\ref{4.5}). Note also that the
statistical gamma-quanta emission given by Eq. (\ref{BW}) does not exist at
high temperatures $T>T_{\mathrm{\lim }}^{(LD)}$\ because of $%
C_{L}^{(LD)}(T)\leq 0 $ and a drop is unstable for this temperature regime.
That means that the dashed line in \textrm{Fig. 3} does not occur for $T>T_{%
\mathrm{\lim }}^{(LD)}$

Some peculiarities of forming of the resonance eigenenergy $E_{\mathrm{GQR}}$
and the corresponding width $\Gamma _{\mathrm{GQR}}$ for the nucleus $^{208}%
\mathrm{Pb}$ are shown in \textrm{Figs. 4} and \textrm{5}. The
eigenfrequencies $\omega $ are derived by the secular equation (\ref{eigen1}%
). In general, the eigenfrequency $\omega $ depends on both the liquid drop
stiffness coefficient $C_{L}^{(LD)}$ and the specific\ one $C_{L}^{\prime
}(\omega )$ caused by the Fermi surface distortions. The dashed line in
\textrm{Fig. 4} represents the result for the classical liquid drop, i.e.,
with $C_{L}^{\prime }(\omega )=0$. The Fermi-liquid eigenfrequencies $\omega
$ (solid lines in \textrm{Fig. 4}) are shifted up with respect to the liquid
drop solution (dashed line) due to the strong enhancement of the stiffness
coefficient caused by the Fermi-surface distortion (\textrm{FSD}) effect. A
shift down of the line \textrm{1} at $T=0$\ for a small value of relaxation
parameter $\beta $ occurs because of a strong hindrance of the \textrm{FSD}
effect in the frequent collision regime. The liquid drop eigenfrequency
(dashed line in \textrm{Fig. 4}) disappears at the limiting temperature $%
T_{0}\approx T_{\mathrm{\lim }}^{(LD)}=7.7\ \,\mathrm{MeV}$, see Eq. (\ref%
{tlim}). An increase of the relaxation parameter $\beta $ provides a
significant contribution $C_{L}^{\prime }$ to the stiffness coefficient
caused by the Fermi-surface distortion effect. Due to this fact the
resonance eigenfrequency $\mathrm{Re}\omega $ exists for temperatures $T_{0}$
higher than the limiting one $T_{\mathrm{\lim }}^{(LD)}$. \ The threshold
for the Fermi-liquid drop eigenfrequencies $\mathrm{Re}\omega $ depends
significantly on the relaxation parameter $\beta $ (see the existence
regions for the curves 1, 2 and 3 in \textrm{Fig. 4}).

\bigskip

For each $\beta $ there are few
solutions to Eq. (\ref{eigen1}). One of them, $\omega ^{(1)}$, is purely
imaginary (see dotted line in \textrm{Fig. 5}). Two of them $\pm \mathrm{Re}%
\,\omega -i\mathrm{Im}\omega $ are located symmetric with respect to the
imaginary axis. Note also that the solution $\omega ^{(1)}$ has the positive
imaginary part for $T_{0}>T_{\mathrm{\lim }}^{(LD)}$ giving rise to an
exponentially growing deformation (unstable mode).\ The motion becomes
overdamped in the temperature regions where $\mathrm{Re}\hbar \omega =0$ (see
the dashed paths in \textrm{Fig. 5}).

\bigskip

\section{Conclusions}

\bigskip

Starting from the collisional Landau-Vlasov kinetic equation with a random
force, we have derived the Langevin-like equation for the surface
fluctuations of the particle distribution in a Fermi system. The main
feature of these fluctuations is that the higher multipole modes ($L\geq 2$)
are strongly influenced by the Fermi-surface distortion effects: the
stiffness coefficient contains an additional contribution $C_{L}^{\prime
}(\omega )$ (see Eq. (\ref{2.20})) and the friction coefficient $\gamma _{L}$%
, Eq. (\ref{2.27}), includes the collisional relaxation phenomena. We have
obtained the random-force correlation function (\ref{2.29}) for the general
case where we also take into account retardation and memory effects in the
relaxation time $\tau (\omega ,T)$. Accounting of the retardation and memory
effects plays an important role in order to obtain a correct transition from
the quantum mechanical regime in cold system to the classical regime at high
temperatures.

The effects of the dynamic distortion of the Fermi surface on the collective
motion lead to the peculiarities of the random-force correlation function
which do not occur in a classical system. The spectral correlation function (%
\ref{2.29}) is independent of $\omega $ and corresponds to a white noise in
the high temperature regime at $\omega \tau \rightarrow 0$ whereas in the
opposite quantum regime at $\omega \tau \rightarrow \infty $ it corresponds
to a blue noise (\ref{4.21}). The behavior of the radiation spectral density
$J_{L}(\omega )$ at different temperatures reflects the above mentioned
peculiarities of the random-force correlation function. We predict a strong
dependence of the shape of the curves $J_{L}(\omega )$ on the retardation
effects ($\omega $-dependence) in the friction coefficient (\ref{2.27}) and,
consequently, on the parameters $\zeta $ and $\beta $ at $L\geq 2$.

Our approach to the shape fluctuations and to the corresponding radiation is
essentially classical. However, due to the Landau's ansatz (\ref{2.14}), the
quantum effects are returned into the fluctuation problem and the
correlation functions (\ref{2.29}) contain contributions from both quantum
and thermal fluctuations. This aspect of the fluctuation theory allows us to
reproduce a standard result (\ref{4.24}) of the quantum theory for the
spectral density of radiation in cold system at zero friction.

Finally, we would like to stress that the fluctuational photon emission,
presented in this paper, does not appear as a new additional source of
radiation but only as a method for determination of radiation which allows
us to include both quantum and thermal emissions of photons in a common
consideration.

\bigskip

\section{Appendix}

\bigskip

Taking three first moments in $\mathbf{p}$-space from Eqs. (\ref{2.1}) and (%
\ref{2.2}) we can derive a closed set of equations for the following moments
of the distribution function, namely, local particle density $\rho $,
velocity field $u_{\nu }$ and pressure tensor $P_{\nu \mu }$, in the form
\begin{equation}
{\frac{\partial }{\partial t}}\delta \rho =-\nabla _{\nu }(\rho _{\mathrm{eq}%
}u_{\nu }),  \label{2.3}
\end{equation}%
\begin{equation}
m\rho _{\mathrm{eq}}{\frac{\partial }{\partial t}}u_{\nu }+\rho _{\mathrm{eq}%
}\nabla _{\nu }\left( {\frac{\delta ^{2}\mathcal{E}}{\delta \rho ^{2}}}%
\right) _{\mathrm{eq}}\delta \rho +\nabla _{\mu }P_{\nu \mu }^{\prime }=0,
\label{2.4}
\end{equation}%
\begin{equation}
{\frac{\partial }{\partial t}}P_{\nu \mu }^{\prime }+P_{\mathrm{eq}}(\nabla
_{\nu }u_{\mu }+\nabla _{\mu }u_{\nu }-{\frac{2}{3}}\delta _{\nu \mu }\nabla
_{\alpha }u_{\alpha })=I_{\nu \mu }+y_{\nu \mu }.  \label{2.5}
\end{equation}%
Here
\begin{equation}
\delta \rho =\int {\frac{gd\mathbf{p}}{(2\pi \hbar )^{3}}\delta }f,\quad {u}%
_{\nu }={\frac{1}{\rho }}\int {\frac{gd\mathbf{p}}{(2\pi \hbar )^{3}}}{\frac{%
{p}_{\nu }}{m}\delta }f  \label{rhou}
\end{equation}%
$g$ is the spin-isospin degeneracy factor, $\mathcal{E}$ is the internal
energy density, which is the sum of the kinetic energy density of the Fermi
motion and the potential energy density associated with the nucleon-nucleon
interaction. The tensor $I_{\nu \mu }$ is the second moment of the collision
integral
\begin{equation}
I_{\nu \mu }={\frac{1}{m}}\int {\frac{d\mathbf{p}}{(2\pi \hbar )^{3}}}p_{\nu
}p_{\mu }\mathrm{St}[f]  \label{2.8}
\end{equation}
and $y_{\nu \mu }$ gives the contribution from the random force
\begin{equation}
y_{\nu \mu }={\frac{1}{m}}\int {\frac{d\mathbf{p}}{(2\pi \hbar )^{3}}}p_{\nu
}p_{\mu }y.  \label{2.9}
\end{equation}%
The equilibrium pressure, $P_{\mathrm{eq}}$, is given by
\begin{equation}
P_{\mathrm{eq}}={\frac{1}{3m}}\int {\frac{d\mathbf{p}}{(2\pi \hbar )^{3}}}%
p^{2}f_{\mathrm{eq}}  \label{2.6}
\end{equation}%
and $P_{\nu \mu }^{\prime }$\ is the deviation of the pressure tensor from
its isotropic part due to the Fermi surface distortion
\begin{equation}
P_{\nu \mu }^{\prime }={\frac{1}{m}}\int {\frac{d\mathbf{p}}{(2\pi \hbar
)^{3}}}(p_{\nu }-mu_{\nu })(p_{\mu }-mu_{\mu }){\delta }f.  \label{pressure}
\end{equation}

Using the Fourier transformation for the pressure
\begin{equation}
P_{\nu \mu }^{\prime }(t)=\int {\frac{d\omega }{2\pi }}e^{-i\omega t}P_{\nu
\mu ,\omega }^{\prime }  \label{2.10}
\end{equation}%
and similarly for the other time dependent variables we find the solution to
Eq. (\ref{2.5}) as
\begin{equation}
P_{\nu \mu ,\omega }^{\prime }={\frac{{i\omega \tau -(\omega \tau )^{2}}}{{%
1+(\omega \tau )^{2}}}}P_{\mathrm{eq}}\Lambda _{\nu \mu ,\omega }+{\frac{%
\tau }{{1+(\omega \tau )^{2}}}}(1+i\omega \tau )y_{\nu \mu ,\omega },
\label{2.11}
\end{equation}%
where we used the symbol
\begin{equation}
\Lambda _{\nu \mu ,\omega }=\nabla _{\nu }\chi _{\mu ,\omega }+\nabla _{\mu
}\chi _{\nu ,\omega }-{\frac{2}{3}}\delta _{\nu \mu }\nabla _{\lambda }\chi
_{\lambda ,\omega }  \label{2.12}
\end{equation}%
for this combination of gradients of the Fourier transform $\chi _{\nu
,\omega }$ of the displacement field. The time derivative of $\chi _{\nu }(%
\mathbf{r},t)$ is defined as the velocity field, hence
\begin{equation}
u_{\nu ,\omega }=-i\omega \chi _{\nu ,\omega }.  \label{2.13}
\end{equation}%
To obtain Eq. (\ref{2.11}) we have also used the fact that the tensor $%
I_{\nu \mu }$, Eq. (\ref{2.8}), can be reduced to
\begin{equation}
I_{\nu \mu ,\omega }=-{\frac{1}{\tau }}P_{\nu \mu ,\omega }^{\prime },
\label{st2}
\end{equation}%
due to our restriction to quadrupole deformation of the Fermi surface.

From Eqs. (\ref{2.3}), (\ref{2.4}) and (\ref{2.11}) we find the equation of
motion for the displacement field $\chi _{\nu ,\omega }$ in the form
\begin{equation}
-\rho _{\mathrm{eq}}\omega ^{2}\chi _{\nu ,\omega }+\hat{\mathcal{L}}\chi
_{\nu ,\omega }=\nabla _{\mu }(\sigma _{\nu \mu ,\omega }+s_{\nu \mu ,\omega
}),  \label{2.16a}
\end{equation}%
where the conservative terms are abbreviated by
\begin{equation}
\hat{\mathcal{L}}\chi _{\nu ,\omega }=-{\frac{1}{m}}\rho _{\mathrm{eq}%
}\nabla _{\nu }\left( {\frac{\delta ^{2}\mathcal{E}}{\delta \rho ^{2}}}%
\right) _{\mathrm{eq}}\nabla _{\mu }\rho _{\mathrm{eq}}\chi _{\mu ,\omega }-%
\mathrm{Im}\left( {\frac{\omega \tau }{{1-i\omega \tau }}}\right) \nabla
_{\mu }{\frac{P_{\mathrm{eq}}}{m}}\Lambda _{\nu \mu ,\omega },  \label{2.17a}
\end{equation}%
$\sigma _{\nu \mu }$ is the viscosity tensor
\begin{equation}
\sigma _{\nu \mu ,\omega }=-i(\omega /m)\eta (\omega )\Lambda _{\nu \mu
,\omega }  \label{2.18}
\end{equation}%
with the viscosity coefficient
\begin{equation}
\eta (\omega )=\mathrm{Re}\left( {\frac{\tau }{{1-i\omega \tau }}}\right) P_{%
\mathrm{eq}}\,,  \label{2.19}
\end{equation}%
and $s_{\nu \mu ,\omega }$ is the random pressure tensor
\begin{equation}
s_{\nu \mu ,\omega }=-{\frac{\tau (1+i\omega \tau )}{m(1+(\omega \tau )^{2})}%
}y_{\nu \mu ,\omega }.  \label{2.20a}
\end{equation}

The correlation properties of $s_{\nu \mu ,\omega }$ can be obtained for the
general case where we also take into account retardation and memory effects
in the system, see Ref. \cite{kiko96} for details. Using the correlation
properties of the random tensor $y_{\nu \mu ,\omega }$ and the
fluctuation-dissipative theorem \cite{lali63}, we find for the ensemble
average of
\begin{equation*}
{\frac{1}{2}}[s_{\nu \mu ,\omega }(\mathbf{r});s_{\nu ^{\prime }\mu ^{\prime
},\omega ^{\prime }}(\mathbf{r^{\prime }})]_{+}={\frac{1}{2}}\left[ s_{\nu
\mu ,\omega }(\mathbf{r})s_{\nu ^{\prime }\mu ^{\prime },\omega ^{\prime }}(%
\mathbf{r^{\prime }})+s_{\nu ^{\prime }\mu ^{\prime },\omega ^{\prime }}(%
\mathbf{r^{\prime }})s_{\nu \mu ,\omega }(\mathbf{r})\right]
\end{equation*}%
the result
\begin{equation*}
{\frac{1}{2}}\overline{[s_{\nu \mu ,\omega }(\mathbf{r});s_{\nu ^{\prime
}\mu ^{\prime },\omega ^{\prime }}(\mathbf{r^{\prime }})]_{+}}
\end{equation*}%
\begin{equation}
={\frac{4\pi }{m^{2}}}E(\omega ,T)\eta (\omega )\delta (\mathbf{r}-\mathbf{%
r^{\prime }})\delta (\omega +\omega ^{\prime })[\delta _{\nu \nu ^{\prime
}}\delta _{\mu \mu ^{\prime }}+\delta _{\nu \mu ^{\prime }}\delta _{\mu \nu
^{\prime }}-{\frac{2}{3}}\delta _{\nu \mu }\delta _{\nu ^{\prime }\mu
^{\prime }}],  \label{2.21a}
\end{equation}%
where
\begin{equation}
E(\omega ,T)={\frac{\hbar \omega }{2}}\coth {\frac{\hbar \omega }{2T}}.
\label{coth}
\end{equation}%
We have preserved the constant $\hbar $ in Eq. (\ref{coth}) in order to
stress the fact that both quantum and thermal fluctuations are involved in
Eq. (\ref{2.21a}).

For the description of small amplitude oscillations of a certain
multipolarity $L$ of a liquid drop we specify the liquid surface as
\begin{equation}
r=R(t)=R_{0}[1+\sum_{M}\alpha _{LM}(t)Y_{LM}(\theta ,\phi )].  \label{rta}
\end{equation}%
We write the displacement field $\chi _{\nu }(\mathbf{r},t)$ for an
incompressible and irrotational flow, $\nabla _{\nu }\chi _{\nu }=0$, as
\cite{nisi80}
\begin{equation}
\chi _{\nu }(\mathbf{r},t)={\sum_{M}a_{LM,\nu }(\mathbf{r})\alpha _{LM}(t)},
\label{chia}
\end{equation}%
where
\begin{equation}
a_{LM,\nu }(\mathbf{r})={\frac{1}{LR_{0}^{L-2}}}\nabla _{\nu
}(r^{L}Y_{LM}(\theta ,\phi )).  \label{ampla}
\end{equation}%
Multiplying Eq. (\ref{2.16a}) by $ma_{LM,\nu }^{\ast }$, summing over $\nu $
and integrating over $\mathbf{r}$-space, we obtain the Langevin equation for
the collective variables,
\begin{equation}
-\omega ^{2}m_{L}\alpha _{LM,\omega }+(C_{L}^{(LD)}+C_{L}^{\prime })\alpha
_{LM,\omega }-i\omega \gamma _{L}(\omega )\alpha _{LM,\omega }=\xi
_{LM,\omega }.  \label{2.28}
\end{equation}%
The collective mass $m_{L}$ is found to be
\begin{equation}
m_{L}=m\int d\mathbf{r}\rho _{\mathrm{eq}}\sum_{\nu }|a_{LM,\nu }|^{2}={%
\frac{3}{4\pi L}}AmR_{0}^{2}.  \label{bla}
\end{equation}%
The static stiffness coefficient $C_{L}^{(LD)}$ is derived from the first
term on the right hand side of Eq. (\ref{2.17a}) and is given by \cite%
{hamy88}
\begin{equation}
C_{L}^{(LD)}={\frac{1}{4\pi }}(L-1)(L+2)b_{S}A^{2/3}-{\frac{5}{2\pi }}{\frac{%
{L-1}}{{2L+1}}}b_{C}{\frac{Z^{2}}{A^{1/3}}}.  \label{2.30}
\end{equation}

The random force $\xi _{LM,\omega }$ in Eq. (\ref{2.28}) is related to the
random pressure tensor $s_{\nu \mu ,\omega }$ by
\begin{equation}
\xi _{LM,\omega }=-m\int d\mathbf{r}\,s_{\nu \mu ,\omega }\nabla _{\mu
}a_{LM,\nu }^{\ast }.  \label{2.36}
\end{equation}%
Using Eqs. (\ref{2.21}) and (\ref{2.25}) we obtain the spectral correlation
function $\overline{(\xi _{LM})_{\omega }^{2}}$ of the random force $\xi
_{LM}(t)$:
\begin{equation}
\overline{(\xi _{LM})_{\omega }^{2}}=2\,E(\omega ,T)\,\eta (\omega )\int d%
\mathbf{r}\,\overline{\Lambda }_{\nu \mu }^{(LM)}\nabla _{\mu }a_{LM,\nu
}^{\ast }=2\,E(\omega ,T)\,\gamma _{L}(\omega ).  \label{2.37}
\end{equation}%
The basic property of the random variable $y$, in Eq. (\ref{2.1}), $%
\overline{y}=\overline{y_{\nu \mu }}=0$ transfers to both, the random
pressure tensor, $\overline{s_{\nu \mu ,\omega }}=0$, and the random force, $%
\overline{\xi _{LM,\omega }}=0$.

\newpage

\begin{center}
\bf{Figure captions}
\end{center}

Fig.~1: The eigenenergies of the isoscalar \textrm{GQR} $(L=2)$ versus 
the nuclear mass number $A$. The
results are obtained from the secular equation (\protect\ref{eigen1}) with 
$\protect\beta =0.8\ $\textrm{MeV}. The experimantal data are taken from 
Ref.\protect\cite{bert81}. The dashed line is for the traditional liquid drop
model (LDM) with $C_{L}^{\prime }(\protect\omega )=0$, Ref. \protect\cite%
{bomo2}.

\bigskip

Fig.~2: The same as in \textrm{Fig. 1} for the widths $\Gamma $ of the
\textrm{GQR}. The dashed lines are for the one body dissipation
(wall-formula \protect\cite{blbo78} and modified wall-formula \protect\cite%
{siko78}).

\bigskip

Fig.~3: The spectral density of the quadrupole gamma-quanta emission $%
J_{L}(\protect\omega )$ for temperatures $T=3\,\mathrm{MeV}<T_{\mathrm{\lim }%
}^{(LD)}$ (curves 1) and $T=8\,\ \mathrm{MeV}>T_{\mathrm{\lim }}^{(LD)}$
(curve 2).\ The solid lines were obtained using Eq. (\protect\ref{4.5})\ for
$\protect\zeta =4\,\protect\pi ^{2}$ (Landau's prescription \protect\cite{lp}%
) and value of relaxation parameter $\protect\beta =0.8\ \mathrm{\,MeV.}$The
dashed line 1 is for the statistical emission of $\protect\gamma $-quanta
given by Eq. (\protect\ref{BW}) which was normalized to the same integral
emission as for solid line 1.

\bigskip

Fig.~4: Dependence of the resonance eigenenergy 
$\hbar\protect\omega _{R}=\mathrm{Re}\hbar \protect\omega $ on the temperature $T$
for three different values of the relaxation parameter $\protect\beta $ (the
solid lines 1, 2 and 3 for $\protect\beta =0.8\ \mathrm{MeV}$, $2.4\ \mathrm{%
MeV}$ and $9.8\ \mathrm{MeV}$, respectively). The dashed line is for the
pure liquid drop regime from Eq. (\protect\ref{2.26}) with $C_{L}^{\prime }(%
\protect\omega )=0$. 

\bigskip

Fig.~5: The same as in \textrm{Fig. 4} but for the value of $\mathrm{Im}%
\hbar \protect\omega $. The dashed paths are the solutions for the regions
where $\mathrm{Re}\hbar \protect\omega =0$.











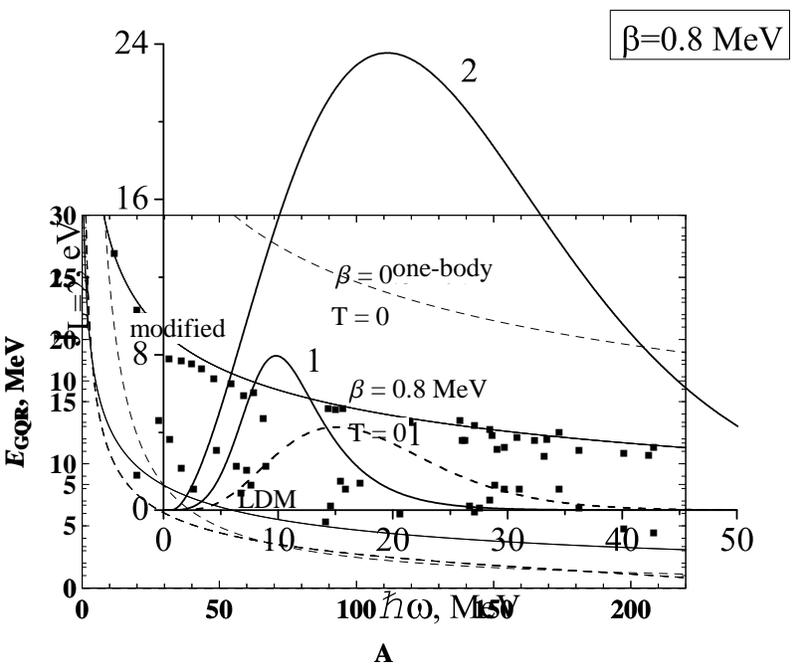
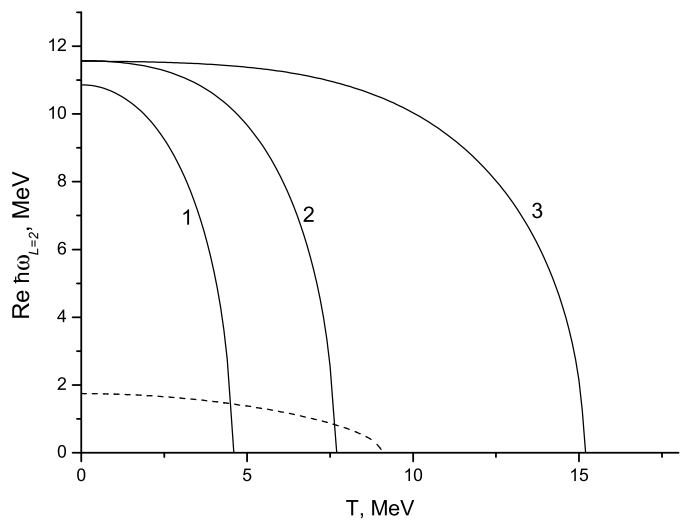
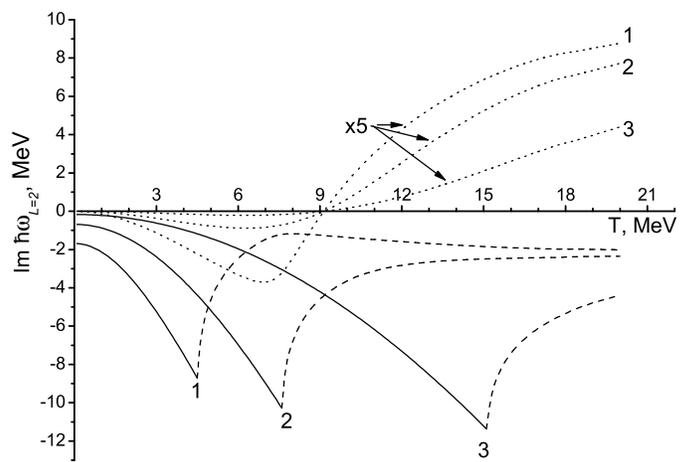
\end{document}